\documentclass[prl,aps,showpacs,tightenlines,twocolumn,nofootinbib]{revtex4}
\usepackage{amsmath}
\usepackage{graphicx}
\usepackage{color}
\usepackage{amssymb}

\newcommand{\order}[1]{\mathcal{O}\!\left(#1\right)}
\newcommand{\VeV}[1]{\left \langle #1 \right \rangle}

\newcommand{\CL}{\mathcal{L}}

\newcommand{\vect}[1]{\boldsymbol{#1}}
\newcommand{\bk}{\vect{k}}

\newcommand{\Mpl}{M_{_\mathrm{P}}}
\newcommand{\Mpc}{\mathrm{Mpc}}
\newcommand{\km}{\mathrm{km}}
\newcommand{\ud}{\mathrm{d}}
\newcommand{\us}{\mathrm{s}}
\newcommand{\ur}{\mathrm{r}}
\newcommand{\um}{\mathrm{m}}
\newcommand{\uinf}{\mathrm{inf}}
\newcommand{\uini}{\mathrm{ini}}

\newcommand{\TeV}{\mathrm{TeV}}
\newcommand{\eV}{\mathrm{eV}}
\newcommand{\rhoinf}{\rho_\uinf}
\newcommand{\Einf}{E_\uinf}
\newcommand{\Hinf}{H_\uinf}
\newcommand{\OmegaL}{\Omega_{_\Lambda}}
\newcommand{\OmegaR}{\Omega_{\ur}}
\newcommand{\OmegaP}{\Omega_\phi}
\newcommand{\wphi}{w_\phi}
\newcommand{\rhorad}{\rho_\ur}
\newcommand{\rhomat}{\rho_\um}
\newcommand{\phiini}{\phi_\uini}
\begin{document}

\title{Dark energy from primordial inflationary quantum fluctuations}

\author{Christophe Ringeval}
\email{christophe.ringeval@uclouvain.be}
\affiliation{Institute of Mathematics and Physics, Centre for
  Cosmology, Particle Physics and Phenomenology, Louvain University, 2
  Chemin du Cyclotron, 1348 Louvain-la-Neuve, Belgium}

\author{Teruaki Suyama}
\email{teruaki.suyama@uclouvain.be}
\affiliation{Institute of Mathematics and Physics, Centre for
  Cosmology, Particle Physics and Phenomenology, Louvain University, 2
  Chemin du Cyclotron, 1348 Louvain-la-Neuve, Belgium}
\affiliation{Research Center for the Early Universe, Graduate School
  of Science, The University of Tokyo, Tokyo 113-0033, Japan}

\author{Tomo Takahashi}
\email{tomot@cc.saga-u.ac.jp}
\affiliation{Department of Physics, Saga University, Saga 840-8502, Japan}

\author{Masahide Yamaguchi}
\email{gucci@phys.titech.ac.jp}
\affiliation{Department of Physics, Tokyo Institute of Technology, Tokyo
152-8551, Japan}

\author{Shuichiro Yokoyama}
\email{shu@a.phys.nagoya-u.ac.jp}
\affiliation{Department of Physics and Astrophysics, Nagoya University,
Aichi 464-8602, Japan}
\date{\today}

\begin{abstract}
  We show that current cosmic acceleration can be explained by an
  almost massless scalar field experiencing quantum fluctuations
  during primordial inflation. Provided its mass does not exceed the
  Hubble parameter today, this field has been frozen during the
  cosmological ages to start dominating the universe only recently. By
  using supernovae data, completed with baryonic acoustic oscillations
  from galaxy surveys and cosmic microwave background anisotropies, we
  infer the energy scale of primordial inflation to be around a few
  TeV, which implies a negligible tensor-to-scalar ratio of the
  primordial fluctuations. Moreover, our model suggests that inflation
  lasted for an extremely long period. Dark energy could therefore be
  a natural consequence of cosmic inflation close to the electroweak
  energy scale.
\end{abstract}

\maketitle

\section{Introduction}

In the past decade, various cosmological observations have accumulated
evidence that the universe is currently undergoing accelerated
expansion~\cite{Perlmutter:1998np, Riess:1998cb, Amanullah:2010vv,
  Tegmark:2006az, Larson:2010gs}. Although cosmic acceleration can be
triggered by a non-vanishing cosmological constant, one still has to
explain why it is so small and why it has started to dominate the
energy content of the universe only
recently~\cite{Zlatev:1998tr}. These questions have motivated the
exploration of alternative explanations all referred to as dark
energy. Among them, the quintessence models consider a scalar field
rolling down a potential in a way similar to the mechanism at work
during primordial inflation~\cite{Guth:1980zm, Linde:1983gd,
  Peebles:1998qn}. A quintessence field yields a time-dependent
equation of state $w(t)=P/\rho>-1$, $P$ and $\rho$ being its pressure
and energy density.  A cosmological constant being equivalent to
$w=-1$, quintessence models generically predict a different expansion
history of the universe, and this can be tested by observations.
According to Ref.~\cite{Caldwell:2005tm}, the quintessence models can
be divided into ``freezing'' and ``thawing'' types.  The former yields
a decreasing $w(t)$ which approaches $-1$ at low redshifts. This type
of model has been intensively used in the literature to address the
coincidence problem. Indeed, assuming the potential to be of the
Ratra--Peebles kind, the quintessence field tracks a cosmological
attractor which erases memory of the initial
conditions~\cite{Ratra:1987rm}. Once the field energy scale is
adjusted to the current dark energy value, its recent domination is
automatic. On the other hand, the thawing type gives $w\simeq -1$ at
high redshifts but can evolve and start deviating from this value at
low redshifts, exactly as during the inflationary graceful exit. These
models have been less explored than the freezing type due to their
dependence on the initial field values. Initial conditions are not
washed out by an attractor mechanism and therefore constitute
additional and unwanted model parameters. In this letter, we show that
inflationary cosmology solves this problem by giving natural initial
values for the quintessence field which can explain the current
acceleration. They are set by the quantum generation of field
fluctuations during inflation. Assuming the accelerated expansion
today to be sourced by the quintessence field, we use the supernovae
data to infer the energy scale of inflation which ends up being at a
few $\TeV$, i.e. close to the electroweak symmetry breaking energy
scale. Such a scenario implies a negligible tensor-to-scalar ratio of
the primordial cosmological perturbations and a reheating temperature
also around a few $\TeV$. Combined with cosmic microwave background
(CMB) bounds, the class of allowed inflationary models is thus
severely constrained~\cite{Martin:2010kz}. Moreover, the total number
of e-folds of inflation has to be extremely large, as it could be in a
self-reproducing inflationary scenario.

Linking primordial inflation and dark energy has been originally
discussed in the context of anthropic selection
effects~\cite{Linde:1986dq, Weinberg:1987dv, Garriga:1999bf,
  Garriga:2003hj}. In these approaches, inflationary quantum
fluctuations are used to randomize the possible dark energy density
values while anthropic selection effects favour the ones compatible
with our own existence. Inflationary stochastic effects in
freezing-type quintessence models have been explored in
Ref.~\cite{Malquarti:2002bh} to determine how inflation influences the
likely initial conditions of the field. As shown in
Ref.~\cite{Martin:2004ba}, they indeed have a tendency to push the
freezing type quintessence field away from the region where the
tracker behavior can efficiently wash out the initial conditions. As a
result, freezing quintessence on the tracker today suggests that
inflation lasted a low number of e-folds.

In our model, the initial value of the thawing quintessence field is
directly determined by the energy scale of inflation and keeps its
initial value until low redshifts. Therefore today's energy density of
the quintessence field is almost equal to its initial energy density,
and hence is directly related to the inflationary energy
scale. Checking the consistency of the model provides us a way to
probe primordial inflation from dark energy observations. This is our
main point and an essential difference from the freezing type
quintessence scenario.

\section{Initial field values from inflation}

Let us consider a (real) free massive scalar field $\varphi$ of mass
$m$ such that $m$ is less than the present Hubble scale
$H_0$. Clearly, since the inflationary Hubble parameter $\Hinf \gg
H_0$, we are in the presence of an almost massless scalar field which
will acquire quantum fluctuations during inflation. By decomposing the
scalar field as a homogeneous mode plus linear perturbations (at the
onset of inflation),
\begin{equation}
  \varphi(\vect x,t) = \phi(t) + \delta\phi(\vect x,t),
\end{equation}
and assuming an almost constant value for $\Hinf$, its Fourier modes
after Hubble exit read~\cite{Grishchuk:1974ny, Bunch:1978yq,
  Starobinsky:1980te, Mukhanov:1981xt, Hawking:1982cz}
\begin{eqnarray}
  \left| \delta \phi_{\bk} \right|^2 \simeq \frac{\Hinf^2}{2 k^3} 
  \left(\dfrac{k}{a \Hinf}\right)^{2 m^2/(3 \Hinf^2)}.
\end{eqnarray}
The comoving wavevector is $\bk$ and $a$ stands for the scale
factor. If inflation starts at $a=a_\us$, after $N=\ln(a/a_\us)$
e-folds, super-Hubble fluctuations induce a real space field variance
given by
\begin{equation}
\label{eq:variance}
\begin{aligned}
  \VeV{\delta\phi^2} &\simeq \int_{a_\us \Hinf }^{a \Hinf}
  \frac{\ud^3 \bk}{ (2 \pi)^3} \left| \delta\phi_{\bk} \right|^2 \\ &
  = \dfrac{3 \Hinf^4}{8\pi^2 m^2} \left[ 1- \exp{\left( - \dfrac{2m^2}{3
          \Hinf^2} N \right)} \right] \rightarrow \dfrac{3 \Hinf^4}{8 \pi^2
    m^2}\,,
\end{aligned}
\end{equation}
where the last limit is reached if inflation lasts long enough for the
exponential term to cancel. This result can be reproduced using the
stochastic inflation formalism~\cite{Starobinsky:1986fx,
  Goncharov:1987ir, Starobinsky:1994bd, Martin:2005ir}. The
homogeneous mode evolves as $\phi = \phi_\us \exp{\left[-N m^2/(3
    \Hinf^2) \right]}$ and becomes completely suppressed compared to
the field fluctuations. The same holds for the mean squared field
derivative (in e-folds) which is suppressed by a factor $m^4/\Hinf^4$
compared to Eq.~(\ref{eq:variance}). For $m \ll \Hinf$ the required
number of e-folds could actually be extremely large. Here, one should
notice that the long-wave fluctuation $\delta\phi$ is almost
homogeneous and determines the typical value of the classical field
$\varphi$. Notice that backreaction effects induced by the field
fluctuations over the expansion rate can produce $\order{1}$
variations in the total number of e-folds. However, one can show that
they induce a correction factor to Eq.~(\ref{eq:variance}) given by
$1+\order{\Hinf^2/\Mpl^2}$ which ends up being completely negligible
as soon as $\Hinf \ll \Mpl$. Since after inflation $H \gg m$, Hubble
damping prevents the field from rolling down the potential until the
time at which $m \simeq H \lesssim H_0$. Before going into a detailed
comparison with observations, let us derive some order of magnitude
results. The present energy density of this quintessence field is
roughly
\begin{equation}
\label{eq:Vnow}
V(\phi) \simeq \dfrac{1}{2} m^2 \langle \delta\phi^2 \rangle
\simeq \dfrac{3 \Hinf^4}{16\pi^2},
\end{equation}
and does not depend on $m$. In order to explain dark energy, one
needs
\begin{equation}
\label{eq:Hinf}
\Hinf \simeq \left(\OmegaL\right)^{1/4} \sqrt{4 \pi H_0 \Mpl}\,,
\end{equation}
where $\Mpl$ stands for the reduced Planck mass and $\OmegaL$ is the
current density parameter associated with a cosmological constant. By
using fiducial values for $H_0$ and $\OmegaL$, one gets $\Hinf \simeq
6 \times 10^{-3}\,\eV$~\cite{Komatsu:2010fb}. The energy scale of
inflation is thus\footnote{However, we should keep in mind that this
  constraint is based on the assumption that the cosmological constant
  is adjusted to zero by some unknown mechanism.}
\begin{equation}
\label{eq:Einf}
\Einf \equiv \rhoinf^{1/4} = \left(3 \Mpl^2 \Hinf^2\right)^{1/4} \simeq 5\,\TeV.
\end{equation}
As a result, in the context of our scenario, we rephrase the question
on the smallness of the dark energy density into the relatively small
energy scale (TeV) of inflation, which may be more tractable to
address and surprisingly close to the electroweak energy scale. Notice
that although the mass $m$ does not appear in Eq.~(\ref{eq:Vnow}) and
(\ref{eq:Einf}), it still determines when the quintessence field
starts rolling down the potential and how the equation of state of
dark energy will deviate from $w=-1$. For $\TeV$ scale inflation, our
model predicts a negligible tensor-to-scalar ratio. Using the above
estimates, with $m \lesssim H_0$, the number of e-folds required to
reach the Bunch--Davies fluctuations is $N \simeq \Hinf^2/m^2 \gtrsim
10^{60}$, which may therefore suggest a self-reproducing inflationary
model~\cite{Vilenkin:1983xq, Linde:1986fd}. In the following, we
assume that $N > \Hinf^2/m^2 >10^{60}$. For such a large $N$, the
homogeneous mode of the scalar field is also suppressed significantly
so that the typical value of the classical field $\varphi$ is
completely determined by the long-wave fluctuation $\delta\phi$.

\section{Observational constraints}
In order to constrain our model from current observations, we have
numerically solved the Einstein and Klein--Gordon equations
\begin{equation}
\begin{aligned}
H^2 & = \frac{1}{3\Mpl^2} \left( \rhorad + \rhomat + \rho_\phi \right), \\
  \ddot{\phi} & +3 H \dot {\phi} + m^2 \phi = 0,
\end{aligned}
\end{equation}
for various values of the parameters $(m,\phiini,H_0)$. The value
$\phiini$ denotes one realisation of the quintessence field
fluctuations in our Hubble patch and $\rhorad$ and $\rhomat$ are the
energy density associated with radiation and matter, respectively. The
integration is started at high redshift, when the dark energy is
sub-dominant, to the time at which $H=H_0$. We have fixed the current
radiation energy density to $\OmegaR h^2 = 4.15 \times 10^{-5}$ and
dropped the decaying mode for the field. By solving the evolution
equations, we obtain $H$ as a function of the redshift $z$ which can
be used to compare the model with observational data.  We have used
type Ia supernovae (SN) redshift-distance modulus relations given in
Ref.~\cite{Amanullah:2010vv} complemented with the redshift-distance
relations at $z=0.2$ and $z=0.35$ coming from the baryon acoustic
oscillations (BAO) data~\cite{Percival:2009xn}. Finally, we have used
the angular scale and height of the first peak of the CMB power
spectrum given in Ref.~\cite{Komatsu:2010fb}. Our likelihood has
therefore been defined as the product of those three, namely by
summing the chi-squared of SN, BAO and CMB data.
\begin{figure}
\includegraphics[width=8.7cm]{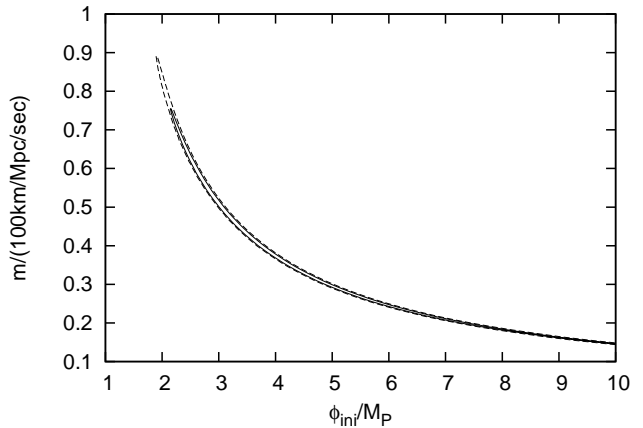}
\caption{One and two-sigma contours of the two-dimensional
  marginalised probability distribution in the plane $(\phiini,m)$
  from SN+BAO+CMB data.}
\label{fig:phi-m}
\end{figure}

In order to test how the model can fit cosmic acceleration, let us
first consider a Jeffreys' prior for the initial field values $\phiini
\ge \Mpl$. Here we are not yet assuming that the initial conditions
are set during inflation, but simply requiring super-Planckian field
values (necessary for triggering acceleration). Together with flat
priors on $m \le 100\,\km/\us/\Mpc$ and $H_0$ around the current
measured value, we have explored the three-dimensional parameter space
with gridding methods. In Fig.~\ref{fig:phi-m}, we have represented
the one and two-sigma contours of the two-dimensional marginalised
probability distribution (over $H_0$) in the plane $(\phiini,m)$. By
marginalising over $\phiini$, we find the mass of the quintessence
field to be constrained by $m < 75 \, \km/\us/\Mpc$ (at $95\%$ of
confidence), which is consistent with our earlier estimation that it
cannot exceed $H_0$ too much.  We also find, as expected, that the
data favour arbitrarily high super-Planckian initial field values. In
order to check our results, we have also derived the field energy
density parameter $\OmegaP$ and its equation of state $\wphi$, both
evaluated at the present time. The contours plotted in
Fig.~\ref{fig:phi-m} end up being centered around the cosmological
constant case $\OmegaP=0.73$, $\wphi=-1$ and are compatible with the
results of Ref.~\cite{Amanullah:2010vv}.
\begin{figure}
\begin{tabular}{c}
  \includegraphics[width=8.6cm]{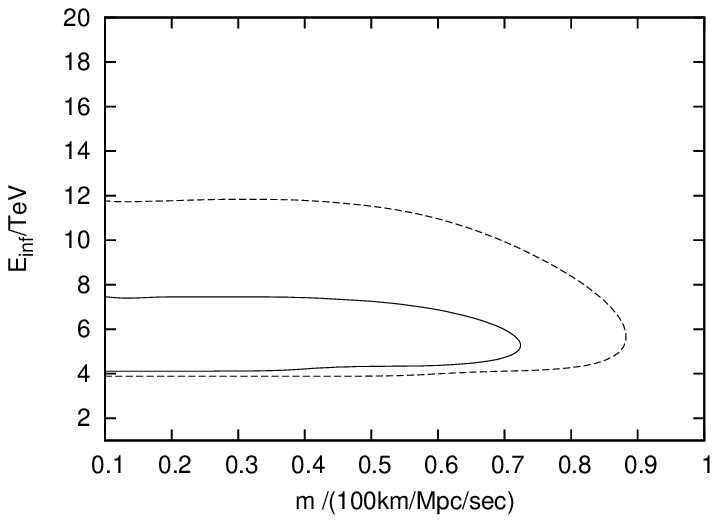} \\
\hspace{0.45cm}  \includegraphics[width=8cm]{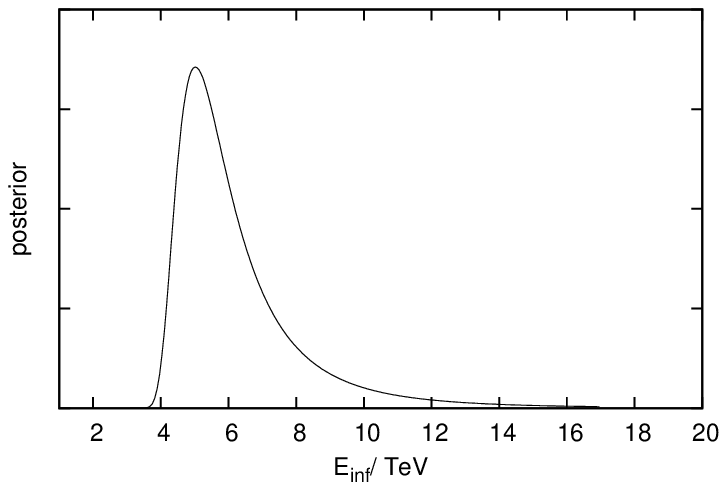}
\end{tabular}
  \caption{One and two-sigma contours in the plane $(m,\Einf)$ from
    SN+BAO+CMB data. The lower panel shows the marginalised posterior
    for $\Einf$. At $95\%$ of confidence, we find $3.8\,\TeV <\Einf<12.1
    \,\TeV$.}
  \label{fig:hinf-m}
\end{figure}

We are now in a position to infer the energy scale of inflation when
the initial field values are set by inflationary quantum
fluctuations. Compared to the above data analysis, our mechanism gives
a prior probability distribution of $\phiini$ which is
Gaussian\footnote{Notice that embedding our mechanism into an explicit
  eternal inflationary model may change this prior, either due to
  volume effects or by a choice of peculiar probability measure. The
  scale factor cut-off measure might however preserve such a
  prior~\cite{Garriga:1999bf, Vanchurin:1999iv, Winitzki:2006rn,
    Linde:2010xz}.} with a standard deviation given by
Eq.~(\ref{eq:variance})
\begin{equation}
  \sigma(\Hinf,m) = \sqrt{\dfrac{3}{8}} \dfrac{\Hinf^2}{\pi m}\,.
\end{equation}
Denoting by $D$ our data sets, and $I$ the prior space, using Bayes'
theorem and marginalising over $\phiini$ and $H_0$ yields
\begin{equation}
\begin{aligned}
  & P(\Hinf,m|D,I) \propto P(\Hinf,m|I) \iint \ud H_0 \dfrac{\ud
    \phiini}{\sqrt{2\pi} \, \sigma} \\ & \times \exp{\left(
    -\dfrac{\phiini^2}{2 \sigma^2} \right)}P(H_0|I)
  \CL(D|m,\phiini,H_0,I) ,
\end{aligned}
\end{equation}
where $\CL$ is the overall SN+BAO+CMB likelihood and $P(\Hinf,m|I)$ the
prior probability distribution on $\Hinf$ and $m$. The scale of
inflation being a priori unknown, we have chosen a flat prior on the
logarithm of $\Hinf$ while the other parameters assume the same prior as
before. In Fig.~\ref{fig:hinf-m}, we have represented the one and
two-sigma confidence intervals of this two-dimensional posterior, up to
the change of variable $\Hinf \rightarrow \Einf$. The lower panel is the
fully marginalised probability distribution for $\Einf$ and the energy
scale of inflation verifies
\begin{equation}
3.8 \, \TeV < \Einf < 12.1 \, \TeV,
\end{equation}
at $95\%$ of confidence. Though this constraint depends on the priors
on $\Hinf$ and $m$, the dependence is small. As discussed below, our
model works for a more general potential, which can slightly change
the constraint. Thus, it is safe to say that our model suggests the
energy scale of inflation $\Einf$ to be around a few TeV for a wide
class of thawing quintessential models.

\section{Discussion}

Although we have focused on a free scalar field so far, the mechanism
discussed here could also be applied to any potential having an
absolute minimum by using the inflationary stochastic
formalism~\cite{Starobinsky:1994bd}. However, the prior probability
distribution for the field $\phiini$ will no longer be Gaussian and
this could therefore shift the favoured values of $\Einf$. Since our
scenario needs an extremely large e-folding number $N \simeq
\Hinf^2/m^2 \gtrsim 10^{60}$, the running of $\Hinf$ along the
inflaton potential must be extremely small. Thus, as a candidate of
such a TeV scale inflation, new inflation might be preferred, which
can indeed realize a self-reproducing era. We also would like to
mention the case that the total number of e-folds is much smaller than
$\Hinf^2/m^2$, in which case Eq.~(\ref{eq:variance}) yields
$\Hinf/\Mpl \propto \sqrt{\OmegaL/N}$. In this limit, we have only a
lower bound $\Einf > \TeV$ and would have to explain why the
homogeneous value $\phi_\us\simeq 0$ after inflation. Notice that CMB
anisotropies imposing $\Hinf/\Mpl < 10^{-5}$, we get an absolute lower
limit for the total number of e-folds $N>10^{9}$, under which the
scale of inflation would actually be too high. Finally, it is worth
stressing that we are in presence of a transient dark energy model. By
rolling down its potential, the thawing quintessence field will
acquire kinetic energy and the current cosmic acceleration will come
to an end.

\begin{acknowledgments}
  We are grateful to Takeshi Chiba, Andr\'e F\"uzfa, Kiyotomo Ichiki,
  Masahiro Kawasaki, Hideo Kodama, J\'er\^ome Martin, Shinji Mukohyama
  and Jun'ichi Yokoyama for enlightening discussions. This work is
  partially supported by the Belgian Federal Office for Scientific,
  Technical, and Cultural Affairs through the Inter-University
  Attraction Pole Grant No. P6/11 (C.~R. and T.~S.); by the
  Grant-in-Aid for Scientific research from the Ministry of Education,
  Science, Sports, and Culture, Japan, No. 19740145 (T.~T.) and
  21740187 (M.~Y.). T.~S. is supported by the JSPS. The work of
  S.~Y. is supported in part by Grant-in-Aid for Scientific Research
  on Priority Areas No. 467 ``Probing the Dark Energy through an
  Extremely Wide and Deep Survey with Subaru Telescope''. He also
  acknowledges support from the Grant-in-Aid for the Global COE
  Program ``Quest for Fundamental Principles in the Universe: from
  Particles to the Solar System and the Cosmos'' from MEXT, Japan.
\end{acknowledgments}

\bibliography{biblio}
\end{document}